\documentclass[conference]{IEEEtran}
\IEEEoverridecommandlockouts
\usepackage{cite}
\usepackage{amsmath,amssymb,amsfonts}
\usepackage{algorithmic}
\usepackage{graphicx}
\usepackage{textcomp}
\usepackage{xcolor}
\def\BibTeX{{\rm B\kern-.05em{\sc i\kern-.025em b}\kern-.08em
    T\kern-.1667em\lower.7ex\hbox{E}\kern-.125emX}}

\usepackage{mathtools}

\usepackage{steinmetz}
\usepackage{authblk}

\usepackage{blindtext}

\begin{document}

\title{Rate-Splitting Multiple Access:\\ A New Frontier for the PHY Layer of 6G\\
\thanks{This work has been partially supported by the U.K. Engineering and Physical Sciences Research Council (EPSRC) under grant EP/R511547/1, and by Huawei Technologies Co., Ltd.}
}

\author[1]{Onur Dizdar}
\author[1]{Yijie Mao}
\author[2]{Wei Han}
\author[1]{Bruno Clerckx}
\affil[1]{Department of Electical and Electronic Engineering, Imperial College London}
\affil[2]{Huawei Technologies, Shanghai, China}
\affil[ ]{Email: \{o.dizdar,y.mao16,b.clerckx\}@imperial.ac.uk, wayne.hanwei@huawei.com}


\maketitle

\begin{abstract}
In order to efficiently cope with the high throughput, reliability,  heterogeneity of Quality-of-Service (QoS), and massive connectivity requirements of future 6G multi-antenna wireless networks, multiple access and multiuser communication system design need to depart from conventional interference management strategies, namely fully treat interference as noise (as commonly used in 4G/5G, MU-MIMO, CoMP, Massive MIMO, millimetre wave MIMO) and fully decode interference (as in Non-Orthogonal Multiple Access, NOMA). This paper is dedicated to the theory and applications of a more general and powerful transmission framework based on Rate-Splitting Multiple Access (RSMA) that splits messages into common and private parts and enables to partially decode interference and treat remaining part of the interference as noise. This enables RSMA to softly bridge and therefore reconcile the two extreme strategies of fully decode interference and treat interference as noise and provide room for spectral efficiency, energy efficiency and QoS enhancements, robustness to imperfect Channel State Information at the Transmitter (CSIT), and complexity reduction. This paper provides an overview of RSMA and its potential to address the requirements of 6G.
\end{abstract}


\section{Introduction}

Rate-Splitting Multiple Access (RSMA) is a multiple access scheme based on the concept of Rate-Splitting (RS) and linear precoding for multi-antenna multi-user communications. RSMA splits user messages into common and private parts, and encodes the common parts into one or several common streams while encoding the private parts into separate streams. The streams are precoded using the available (perfect or imperfect) Channel State Information at the Transmitter (CSIT), superposed and transmitted via the Multi-Input Multi-Output (MIMO) or Multi-Input Single-Output (MISO) channel \cite{clerckx_2016}. All the receivers then decode the common stream(s), perform Successive Interference Cancellation (SIC) and then decode their respective private streams. Each receiver reconstructs its original message from the part of its message embedded in the common stream(s) and its intended private stream.

The key benefit of RSMA is to flexibly manage interference by allowing the interference to be partially decoded and partially treated as noise. RSMA has been demonstrated to embrace and outperform existing multiple access schemes, i.e., Space Division Multiple Access (SDMA), Non-Orthogonal Multiple Access (NOMA), Orthogonal Multiple Access (OMA) and multicasting. Thus, RSMA is a promising enabling technology for 5G and beyond \cite{joudeh_2016_2,clerckx_2019}.

With the ongoing deployment of 5G New Radio (NR), conceptual studies on 6G have already started.  New enabling technologies are evaluated in order to address the enhanced requirements of 6G. In this paper, we discuss these requirements in light of the studies on RSMA and how RSMA can fit as an enabling technology in 6G. 
 

\textit{Notations:} Vectors are denoted by bold lowercase letters. 
Operations $|.|$ and $||.||$ denote the absolute value of a scalar and $l_{2}$-norm of a vector. 
$\mathbf{a}^{H}$ denotes the Hermitian transpose of a vector $\mathbf{a}$. 
$\mathcal{CN}(0,\sigma^{2})$ denotes the Circularly Symmetric Complex Gaussian distribution with zero mean and variance $\sigma^{2}$. 
$\mathbf{I}$ denotes the identity matrix.


\section{Rate-Splitting Multiple Access}
RSMA has been studied in several forms, such as splitting each or some users' messages, or using one or more layers of common messages for $K$ users. Due to lack of space, we give the basics for RSMA based on 1-layer RS for the MISO Broadcast Channel (BC) and refer the interested reader to \cite{mao_2018} for a more comprehensive study.

\begin{figure*}[htbp]
	\centerline{\includegraphics[width=6.7in,height=6.7in,keepaspectratio]{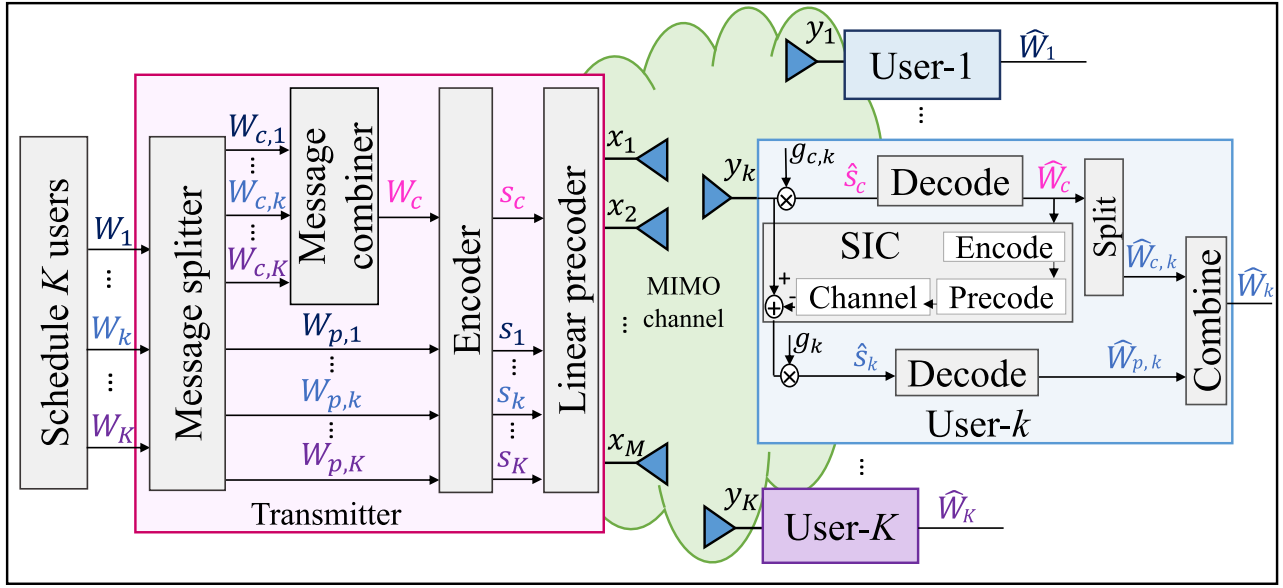}}
	\vspace{-0.2cm}
	\caption{Transmission model of $K$-user 1-layer RSMA}
	\label{fig:system}
	\vspace{-0.6cm}
\end{figure*}

We consider the MISO BC setting consisting of one transmitter with $M$ transmit antennas and $K$ single-antenna users. Figure~\ref{fig:system} illustrates the transmission model of 1-layer RS. The message intended for user $k$, $W_{k}$, is split into common and private parts, i.e., $W_{c,k}$ and $W_{p,k}$, for $k \in \mathcal{K}=\{1,\ldots,K\}$. The split messages are assumed to be independent. The common parts of the messages, $W_{c,k}$, $k\in \mathcal{K}$, are combined into the common message $W_{c}$. The single common message $W_{c}$ and the $K$ private messages $W_{p,k}, k\in\mathcal{K}$ are independently encoded into streams $s_{c}$, $s_{k}$, $k\in\mathcal{K}$, respectively. Linear precoding is applied to all streams. The transmit signal is expressed as
\begin{align}
\mathbf{x}=\mathbf{p}_{c}s_{c}+\sum_{k \in \mathcal{K}}\mathbf{p}_{k}s_{k}.
\end{align}
The transmitted streams satisfy \mbox{$\mathbb{E}\left\lbrace \mathbf{s}\mathbf{s}^{H}\right\rbrace =\mathbf{I}$} where  \mbox{$\mathbf{s}=\left[s_{c}, s_{1}, \ldots, s_{k}\right]$}. 
The average transmit power constraint is $P_{c}+\sum_{k\in\mathcal{K}}P_{k} \leq P$ for $P_{c}=||\mathbf{p}_{c}||^{2}$ and $P_{k}=||\mathbf{p}_{k}||^{2}$.
The signal received by user-$k$ is written as
\begin{align}
	\mathbf{y}_{k}=\mathbf{h}_{k}^{H}\mathbf{x}+n_{k}, \quad k\in\mathcal{K},  
\end{align}
where $\mathbf{h}_{k} \in \mathbb{C}^{M}$ is the channel vector between the transmitter and user-$k$ and $n_{k} \sim \mathcal{CN}(0,\sigma_{k}^{2})$ is the Additive White Gaussian Noise (AWGN) component for user-$k$. 
The receivers apply SIC to detect the common and their corresponding private streams. The common stream is detected first to obtain the common message estimate $\hat{W}_{c}$ by treating the private streams as noise. 
The Signal-to-Interference-plus-Noise Ratio (SINR) for the common stream at user-$k$ is
\begin{align}
	\gamma_{c,k}=\frac{|\mathbf{h}_{k}^{H}\mathbf{p}_{c}|^{2}}{\sum_{i \in \mathcal{K}}|\mathbf{h}_{k}^{H}\mathbf{p}_{i}|^{2}+\sigma_{k}^{2}}.
\end{align}

The maximum rate for the common stream is $  R_{c}\hspace{-0.1cm}=\min\left\lbrace\log_{2}(1+\gamma_{c,1}), \ldots, \log_{2}(1+\gamma_{c,K}) \right\rbrace$.
The common stream consists of the common messages intended for all users, i.e, $R_{c}=\sum_{i \in \mathcal{K}}C_{k}$,
where $C_{k}$ is the portion of the rate of the common stream intended for user-$k$.  

The common stream is reconstructed using $\hat{W}_{c}$ and subtracted from the received signal. The remaining signal is used to detect the private messages $\hat{W}_{p,k}$. 
The SINR and maximum rate for the private stream at user-$k$ is given by
\begin{align}
	\gamma_{k}=\frac{|\mathbf{h}_{k}^{H}\mathbf{p}_{k}|^{2}}{\sum_{i \in \mathcal{K}, i \neq k}|\mathbf{h}_{k}^{H}\mathbf{p}_{i}|^{2}+\sigma_{k}^{2}},
\end{align}
and $R_{k}=\log_{2}(1+\gamma_{k})$, respectively. 
The overall maximum rate for user-$k$ is $R_{k,\mathrm{total}}=C_{k}+R_{k}$.

\section{Communications Beyond 5G: 6G and RSMA}
It is commonly agreed that the Key Performance Indicators (KPIs) of NR, namely, high data rate, ultra-reliable and low-latency communications, and massive connectivity, are also valid for 6G, but with enhanced requirements. More specifically, it is predicted that the throughput requirements will surpass $1$~Tbps, the latency requirement will be below $0.1$~ms and the reliability requirement will be down to a Block Error Rate (BLER) of $10^{-9}$. 

It is almost certain that the abovementioned KPIs are not enough by themselves to describe 6G. Additional and equally important KPIs are predicted to be secrecy, high positioning accuracy ($<10$~cm), low jitter, high energy efficiency ($1$~pJ/b), increased spectral efficiency ($100$~bps/Hz), increased density ($100$ devices per $\mathrm{m}^3$), increased coverage and service availability and increased area traffic capacity ($1$-$10$~Gb/s/$\mathrm{m}^3$) \cite{2,4,5,10}. 

6G targets vertical industries such as health, manufacturing, transport, coverage, and energy. Autonomous vehicles, health applications, wireless factory automation will require connectivity of a massive number of devices and their operation will be critically dependent on instant and reliable information \cite{2,4}. Inter-communication among cars is mandatory to prevent accidents, which implies highly reliable and low-latency communications under mobility \cite{5}. In factory automation, a massive network of robots will require high rate, reliable and low-latency communications.

In addition to the KPIs specified above, several challenging characteristics related to the 6G use cases arise. For example, connectivity and communications of a massive number of devices under mobility will be one critical aspect of 6G. The supported user speed is expected to increase up to $1000$~km/h. In order to increase the network capacity to support dense communications, enabling technologies with spectral efficiencies larger than $100$~b/s/Hz will be required \cite{2}. 

\subsection{RSMA for KPIs and Characteristics of 6G}
We discuss RSMA in light of the predicted KPIs and characteristics of 6G. 
In NR, three core services are defined to address the requirements of the use cases, namely enhanced Mobile Broadband (eMBB), Ultra-Reliable Low-Latency Communications (URLLC) and Massive Machine Type Communications (mMTC). eMBB is the core service for data-driven use cases with a rate requirement of $20$~Gbps in the DL.
URLLC is aimed to provide reliable communications (BLER of $10^{-5}$) with low latency ($1$ ms). 
mMTC is a core service to connect a massive number of low-complexity low-power devices with grant-free access in the uplink (UL).

In order to satisfy the more complicated demands of vertical industries and use cases in 6G, four kinds of core services are identified as combinations of those in NR, namely enhanced eMBB + URLLC, enhanced eMBB + mMTC, \mbox{enhanced URLLC + mMTC}, and tradeoff based \mbox{enhanced eMBB + URLLC + mMTC} \cite{10}. Hybrid core services imply that the task of multiple access becomes even more challenging with the additional KPIs in 6G. 

In this section, we discuss how RSMA can be a unified solution to address the requirements and challenges of the hybrid core services of 6G.  
\subsubsection{Data Rate}
\label{sec:embb}
The data rate requirement in 6G is expected to be on the order of Tbps, much higher than that of NR. In order to be able to support such high rates, first, the problems affecting the performance of the data service in NR should be addressed in 6G.
Field test results for the data service in NR have been reported in \cite{wang_2017, sakai_2020}. A reported problem is multi-user interference due to imperfect CSIT in MIMO beamforming. In \cite{wang_2017}, the problem of outdated CSIT due to mobility is highlighted. The processing delay in the network can be larger than the coherence time of the channel even when the users are moving with a speed of $30$km/h, making the CSIT unusable for multiuser beamforming. 

Another reported source of interference is due to blockage of users in multi-user transmission. In a scenario where one of the multiple users served in the downlink (DL) goes into outage, the Channel State Information (CSI) report from that user cannot be updated, resulting in a rate loss not only for the specific user but also other users in the DL transmission \cite{sakai_2020}. 

Being a multiple access scheme designed for BC with multiple transmit (and receive) antennas in BC, RSMA is naturally a promising candidate for operating under interference in multi-antenna settings. Indeed, numerous studies \cite{dai_2016, papazafeiropoulos_2017, joudeh_2016, joudeh_2016_2, hao_2015, piovano_2016} have demonstrated that RSMA has superior performance compared to SDMA, NOMA and OMA under both perfect and imperfect CSIT. Several causes of imperfect CSIT can be listed as channel estimation errors, finite rate CSI feedback, latency and mobility, subband processing, or phase noise. A theoretical foundation for the superiority of RSMA is the Degrees-of-Freedom (DoF) analysis performed in \cite{joudeh_2016_2} for the specific case of unbounded Gaussian distributed channel estimation errors. The authors assume an estimation error which scales with Signal-to-Noise Ratio (SNR) according to a quality scaling factor, i.e., $\sigma_{e}^{2} \propto O(SNR^{-\alpha})$, where $\sigma_{e}^{2}$ is the error variance, $SNR=P/\sigma^{2}$ ($\sigma^2=\sigma_k^2$, $\forall k\in\mathcal{K}$) and $\alpha \in \left[0,1\right] $ is the quality scaling factor. Denoting the ergodic sum-rates of SDMA and RSMA as $R_{\mathrm{SDMA}}$ and $R_{\mathrm{RSMA}}$, the authors demonstrated that the DoF of SDMA and RSMA are given by
\begin{subequations}
	\begin{align}
	\lim_{P\rightarrow\infty}\frac{R_{\mathrm{SDMA}}(P)}{\log_{2}P}=\max\left\lbrace 1,K\alpha\right\rbrace,  \label{eqn:dof_sdma}\\
	\lim_{P\rightarrow\infty}\frac{R_{\mathrm{RSMA}}(P)}{\log_{2}P}=1+(K-1)\alpha. \label{eqn:dof_rsma}
	\end{align}
\end{subequations}

An implication of the expressions \eqref{eqn:dof_sdma} and \eqref{eqn:dof_rsma} is that when the channel estimation quality drops below a certain threshold $(\alpha < 1/K)$, it is preferred for SDMA to switch to single-user transmission, hence achieving a DoF of $1$. In case such switching is not performed, the DoF is reduced below $1$. For channel estimation errors non-scaling with SNR ($\alpha=0$), such as errors due to mobility and processing delay, or finite rate quantized feedback, the DoF of SDMA (assuming it does not switch to single-user transmission) becomes zero. On the other hand, RSMA achieves a strictly larger DoF than SDMA for all $\alpha \in (0,1)$. Moreover, it achieves a DoF of $1$ for channel errors non-scaling with SNR, implying it is robust to different kinds of CSIT imperfections, and therefore latency and mobility.

The robustness and flexibility of RSMA under heterogenous traffic (unicast, multicast, broadcast) demands and user capabilities, underloaded and overloaded scenarios and different propagation conditions (such as a wide range of disparities of channel strengths and channel directions among users) are demonstrated in \cite{piovano_2016,mao_2018,clerckx_2019,mao_2019_2}. This, combined with the robustness to imperfect CSIT, would also address the rate loss due to interference from mobility and blockage.

A less trivial cause of performance degradation in multiple access is the separate implementation of the Physical (PHY)-layer design (such as precoder design) and the scheduling algorithm of the multiple access scheme separately (in the higher layers), in an effort to reduce the implementation complexity. This suboptimality creates a rate loss especially in the multiple access schemes in which the performance heavily depends on the channel disparities and orthogonalities of the users being served, such as SDMA and NOMA. Therefore, these schemes highly rely on the user topologies and the success of the user grouping and scheduling algorithms. The effects of user grouping and scheduling has been studied in	\cite{dai_2016, mao_2018, mao_2019_2}, in which RSMA has been shown to be more robust to the interference resulting from suboptimal grouping and can achieve a high sum-rate even with very simple schedulers.

In addition to addressing the problems in NR, major changes in the operating spectrum and the transmission strategies are required to provide a data rate of up to $1$~Tb/s/Hz. In order to support the increased bandwidth, frequencies in the range of several hundred GHz or THz have to be used \cite{2,4,5,10}. Even though the communication ranges at those frequency bands are relatively short, the problems and their solutions discussed above can be valid for certain scenarios. 
	
Achieving such high data rates will also require modulations with very high modulation orders, which are highly sensitive to hardware impairments, such as phase noise \cite{4}. RSMA has been demonstrated in \cite{papazafeiropoulos_2017} to be robust for multiuser MIMO in the presence of phase noise and provide a non-saturating sum-rate contrary to SDMA.
Additionally, technologies with higher spectral efficiencies will be sought to achieve such a high rate transmission \cite{2,4,5}. The sum-rate performance of RSMA has been demonstrated to surpass those of SDMA and NOMA under perfect and imperfect CSIT in \cite{piovano_2016,hao_2015,dai_2016,papazafeiropoulos_2017,dai_2017,flores_2020, joudeh_2016_2}. Thus, RSMA is a promising candidate for high data rate services in 6G. 
\subsubsection{Latency and Reliability}
\label{sec:urllc}
As in NR, latency and reliability are key KPIs in 6G, but with enhanced requirements.
Example use cases of services providing reliable and low-latency communications in 6G are intelligent transportation and industrial automation.
In intelligent transportation and fully automated driving, cooperation among the cars for collision avoidance, environmental awareness and high density platooning require strict constraints on reliability and latency. 
In industrial automation, use cases such as factory, process, and power system
automation require very low latency and high reliability \cite{durisi_2016, chen_2018}. The most critical sources of latency are identified as the link establishment, retransmissions caused by channel estimation errors and congestion, and the minimum data block length \cite{chen_2018}. 



An important limitation for the use of multiple antennas in latency critical applications is beamforming based on the channel state. For Frequency-Division Duplex (FDD) and Time-Division Duplex (TDD) systems, obtaining instantaneous CSIT imposes a huge latency factor. In order to benefit from the multiple antenna transmission without instantaneous CSIT, beamforming can be performed based on the second-order statistics (covariance matrix) of the channel. However, this approach decreases the precision of the beam and causes interference in the DL \cite{popovski_2018}. 

As discussed in Section~\ref{sec:embb}, RSMA is a multiple access scheme designed for multi-antenna BC with imperfect CSIT and its performance has been demonstrated to be superior to SDMA, NOMA, and OMA in numerous studies. Thus, RSMA is a promising candidate to achieve a higher performance in low-latency services. More specifically, preliminary results showing the benefits of RSMA using second-order statistics are available in \cite{dai_2016,dai_2017}.  

Retransmissions constitute a large portion of transmission delay. Thus, a key point to reduce latency is to increase robustness, so that no retransmissions are required. It is generally not possible to control the environment and mobility, which may result in interference and packet loss. This is critical especially in low-latency scenarios with multiple access, such as intelligent transportation, where packet loss may result in serious consequences. Since it is not possible to predict interference, a robust scheme is required for connectivity. As expressed above, RSMA has been demonstrated to achieve better performance than other multiple access schemes under interference, thus can be employed to address the retransmission problem in low-latency scenarios.

Reducing the packet size in transmissions is mandatory for the low-latency services. The increased spectral efficiency under perfect and imperfect CSIT hints that RSMA is also a good candidate for transmission with short channel codes \cite{hao_2015,dai_2016,papazafeiropoulos_2017,dai_2017,flores_2020, joudeh_2016_2,piovano_2016}. Preliminary results demonstrating the benefits of RSMA over SDMA and NOMA with finite block length coding can be found in \cite{dizdar_2020}.
\subsubsection{Mobility}
As a major problem in NR, mobility will be encountered more frequently and severely in the use cases of 6G. 
In addition to the broadband data services, Vehicle-to-Everything (V2X) technology is especially prone to the distruptive effects of mobility. 
It is also crucial for communications among cars to be reliable and of low-latency under a wide range of speeds (pedestrian to very high speed) \cite{2,5}. The discussions in Sections~\ref{sec:embb} and \ref{sec:urllc} highlight that RSMA is a robust technique which can deal with interference due to mobility and feedback delay in multi-user MIMO and massive MIMO. The robustness of RSMA makes it a strong enabling technology for use cases with reliability requirements under mobility.  
\subsubsection{Network Density and Dynamic Topology}
From the perspective of access point, one challenge of supporting a massive number of users stems from the amount of feedback in the UL to create a sufficiently high spectral efficiency gain in the DL. As mentioned in the previous section, RSMA has been demonstrated to achieve better performance than SDMA and NOMA under imperfect CSIT. A more specific study on the finite rate CSI feedback \cite{hao_2015} shows that RSMA enables a quantized CSIT feedback overhead reduction over SDMA with Zero Forcing Beamforming (ZFBF). The results encourage the use of RSMA to improve network efficiency.  

Another challenge in the DL is to establish an efficient multiple access transmission without the necessity of employing receivers with high complexity  \cite{bockelmann_2016}. RSMA based on 1-layer RS has been demonstrated to achieve significantly higher sum-rate than NOMA in overloaded scenarios \cite{mao_2018}. It should be noted that 1-layer RS operates with 1 layer of SIC while NOMA requires $K-1$ layers of SIC, implying a large performance gain coupled with a significant complexity reduction. 

The densification of networks, dynamic cell structure, multiple connection options for cell nodes, and mobility result in severe interference in the network \cite{2}. As a result, the problem of managing the inter-cell and intra-cell interference with decreasing cell sizes and increasing resource sharing becomes more complex.
What is equally important as managing the interference is the robustness of the enabling technologies under interference. Furthermore, serving multiple users simultaneously in the same resource blocks with a high spectral efficiency can improve communications under high network density \cite{10}. The sum-rate of RSMA under interference is known to be superior to other multiple access schemes, as also discussed in Section~\ref{sec:embb}. Thus, RSMA is a perfect fit for highly-dense and dynamic networks.
\subsubsection{Energy Efficiency}
Supporting massive connectivity with extreme KPIs and characteristics calls for energy efficient solutions in 6G. RSMA has been shown to be energy-efficient in \cite{mao_2018_2, mao_2019_2, mao_2020} for $2$-user and $K$-user MISO BC with different channel directions and channel strengths, and \cite{rahmati_2019} in UAV networks with fixed and perfectly known locations. 
\subsection{RSMA for Enabling Technologies in 6G}
The KPIs and use cases of 6G require novel enabling technologies as well as enhanced versions of the existing ones. In this section, we discuss such technologies with RSMA. 
\subsubsection{RSMA for Massive MIMO}
The number of antennas at the BS is expected to increase in 6G. As explained in Section~\ref{sec:embb}, massive MIMO is vulnerable to CSIT imperfections, which may occur due to numerous reasons. 

Although many studies on RSMA in imperfect CSIT are valid for massive MIMO, RSMA has been studied specifically for massive MIMO with imperfect CSIT in \cite{dai_2016}. Two-stage beamforming and 2-layer RS is performed to cope with inter-group and intra-group interference. It is demonstrated that the sum-rate achieved by this structure surpasses that of conventional two-stage beamforming for SDMA. Additonally, RSMA has been shown to mitigate hardware impairments and pilot contamination in massive MIMO \cite{papazafeiropoulos_2017, thomas_2020}.

The increase in the number of transmit antennas in turn leads to the need for a larger number of Radio Frequency (RF) chains, which increases the cost of system design. A solution is to use hybrid analog/digital beamforming to reduce the number of RF chains. However, this solution brings forth a complicated feedback procedure in the UL for both analog and digital precoders, both of which require feedback from the transmitters.
An RSMA scheme is investigated in \cite{dai_2017} with one-stage feedback used for analog beamforming and statistical CSIT for digital beamforming. For a given amount of feedback for first stage analog beamforming, it is shown that RSMA achieves comparable sum-rate to that of two-stage feedback, which enables a significant saving in the channel training and feedback for the digital beamforming. 
\subsubsection{RSMA for UAV-aided Communications}
\label{sec:uav}
An interesting topic in the development of NR is the use of Unmanned Aerial Vehicles (UAVs) as BS to provide coverage in areas with no infrastructures or whenever the existing infrastructures are heavily loaded due to disasters or events. 

The increased coverage requirements in 6G call for additional infrastructures to support the network coverage and connectivity, such as aerial networks, which make use of UAVs as BS to provide coverage. 
In addition to the coverage, cell-free networks with mobile UAV BSs may achieve reduced latency compared to fixed infrastructures \cite{2,4,5,10}. UAVs enable a dynamic and flexible network structure that can answer the densification demands efficiently \cite{9}. The idea of using UAVs for network connectivity has been considered in the scope of NR. However, it is expected to reach its full potential in 6G. 

RSMA in UAV-aided communications has been considered for DL mmWave transmission in \cite{rahmati_2019}. For the scenario in which UAVs are served by a terrestrial BS, RSMA is shown to perform superior to NOMA in terms of energy efficiency. The use of UAVs to increase the coverage of Cloud-Radio Access Networks (C-RANs) is considered in \cite{ahmad_2019}, where UAVs replace broken BSs. It is demonstrated that the maximum sum-rate improves when RSMA is applied at the UAVs and BSs.
\subsubsection{RSMA for Satellite Communications}
In addition to supporting the 6G network, satellite networks can be used to connect the space, air, ground, sea and remote areas into the 6G core \cite{10}. 
RSMA for satellite communications has been studied in \cite{yin_2020} with multiple co-channel beams and multiple users within each beam. The authors show that RSMA manages the inter-beam interference efficiently in overloaded multibeam satellite communication scenarios with perfect and imperfect CSIT. 
\subsubsection{RSMA for Mobile Caching}
An emerging method to reduce the latency under increasing data traffic is to cache the contents closer to the users. By enabling content reuse, mobile caching reduces delay and improves the backhaul efficiency \cite{chen_2018,10}. RSMA has been considered for caching in \cite{piovano_2017_2, piovano_2019} by splitting the user messages into cached and uncached parts and transmitting the cached parts in one multicasting stream. It has been demonstrated that RSMA achieves better sum-rate and DoF compared to time-sharing and a latency improvement can be achieved under limited CSIT quality. 
\subsubsection{RSMA for Wireless Powered Communications}
Very low-power communications is especially required in Internet-of-Things (IoT) nodes with non-replaceable batteries in 6G \cite{4}. An approach to achieve a long lifetime for such low-complexity devices is Wireless Power Transfer (WPT), which becomes more feasible in 6G due to dense networks and shorter communication distances \cite{2}. Simultaneous Wireless Information and Power Transfer (SWIPT) performs information and energy transmission to Information Receivers (IRs) and Energy Receivers (ERs) concurrently \cite{clerckx_2019_2}. Combined with energy-efficient communication technologies, SWIPT can be the solution for an overall low-power network solution. 

RSMA with SWIPT has been studied in \cite{mao_2019_3} with separate IRs and ERs. The Weighted Sum Rate (WSR) of IRs is maximized under total transmit power and sum energy constraints for ERs in precoder design. It is demonstrated that the rate region of IRs in RS-assisted SWIPT always spans those of SDMA and NOMA-assisted SWIPT.
\subsubsection{RSMA for Intelligent Reflecting Surfaces}
In order to meet the high energy and spectral efficiency demands, altering the communications environment by means of smart structures is a novel solution. Intelligent Reflecting Surfaces (IRSs) offer the ability to control the signal reflections and refractions in the environment to customize the received signal according to the needs. IRS can be used to enhance the coverage area of small cells and improve communications reliability, making it a tempting technology for 6G \cite{2,4,5,10}. 

A challenge in IRS design is the acquisition of CSI due to the lack of RF chains. This causes imperfect CSIT and thus interference in the communications channel. RSMA is a strong candidate to be used in systems employing IRS since it is robust to imperfect CSIT and can cope with the challenges in IRS better than other technologies.  
\subsubsection{RSMA for Joint Radar/Sensing and Communications}
Intelligent vehicles are expected to employ radar technologies for environemntal awareness and interaction with other devices \cite{2}. This creates a necessity to operate the radar and communications tasks jointly in an efficient manner. In \cite{xu_2020}, beamformer design for a joint radar-communications system with RSMA has been studied. By jointly maximizing the user WSR and minimizing the approximation error for the desired radar beampatterns, RSMA has been shown to achieve a better trade-off between WSR and approximation error than SDMA.
\subsubsection{RSMA for Visible Light Communications}
The data rate demand increasing beyond 1Tb/s brings the need to exploit higher frequency bands. THz and Visible Light Communications (VLC) are two enabling technologies to achieve the high rate demand in short distances and indoors \cite{2,4,5,10}. Application of RSMA in VLC with high SNR levels is promising due to the improved rate performance of RSMA.  
\subsubsection{RSMA for Cooperative Communications}
Power consumption for achieving high data rates over long distances can be effectively reduced by cooperative communications using relays. RSMA has been investigated for cooperative communications for $2$ users and $K$ users in \cite{zhang_2019} and \cite{mao_2019_4}, respectively. Cooperative RS is implemented by letting the relays to forward the decoded common stream to enhance its performance. The advantages of 1-layer RS for $2$ users in terms of rate region have been demonstrated in \cite{zhang_2019}. The results show that RSMA is a promising candidate for cooperative communications and worth investigating in further studies. 
\subsubsection{RSMA for Cognitive Radio}
6G is likely to see the use of Cognitive Radios (CRs) to achieve high connection density under coexistence of different networks. CR may provide a solution to the increase of heterogenous nodes and services in the network \cite{2,10}. RSMA is a natural candidate for use in CR networks as it is designed to operate under interference. 
\begin{figure}[t!]
	\vspace{-0.4cm}
	\centerline{\includegraphics[width=3.8in,height=3.8in,keepaspectratio]{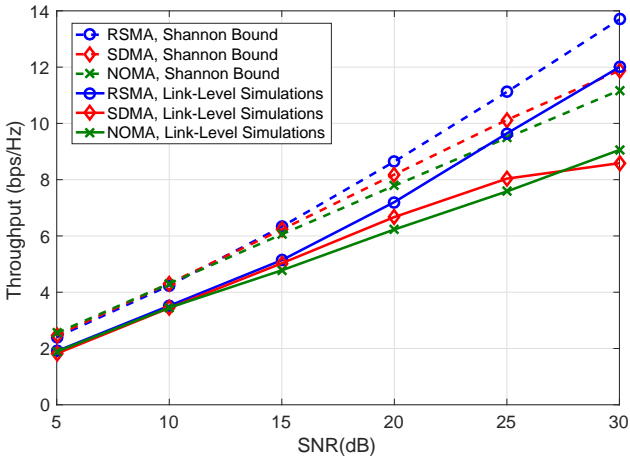}}
	\caption{SNR vs. Throughput, $\alpha=0.6$, No QoS constraint}
	\label{fig:tp_alpha06_noQoS}
	\vspace{-0.55cm}
\end{figure}
\section{Link-Level Simulation Results}
\label{sec:results}
In this section, we demonstrate the performance improvements achieved by RSMA over SDMA and NOMA by Link-Level Simulations (LLS). The details of the transceiver architecture used in the simulations are given in \cite{dizdar_2020}. The precoders for the common and private streams are optimized to maximize the sum-rate of the system. We employ the modulation schemes $4$-QAM, $16$-QAM, $64$-QAM and $256$-QAM and polar coding \cite{arikan_2009} for transmission. The Adaptive Modulation and Coding (AMC) algorithm detailed in \cite{dizdar_2020} selects a suitable modulation-coding rate pair to transmit data.

\begin{figure}[t!]
	\vspace{-0.5cm}
	\centerline{\includegraphics[width=3.8in,height=3.8in,keepaspectratio]{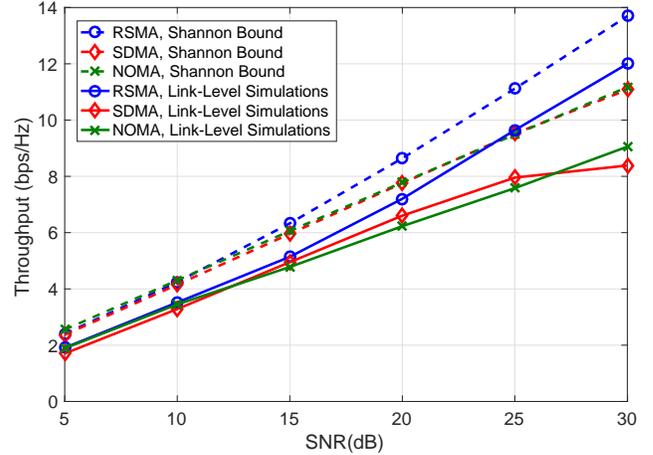}}
	\caption{SNR vs. Throughput, $\alpha=0.6$, $R_{0}=0.1$bps/Hz}
	\label{fig:tp_alpha06_QoS}
	\vspace{-0.55cm}
\end{figure}

We consider a 2-user MISO BC scenario, in which a central node employing $n_{t}=2$ antennas aims to communicate with 2 users, each employing a single antenna.
The transmitter does not have access to instantanous perfect CSI. We assume a block fading channel model $\mathbf{H}=[\mathbf{h}_{1}, \mathbf{h}_{2}]$ with channel estimation errors 
\begin{align}
\mathbf{H}=\sqrt{1-\sigma_{e}^{2}}\widehat{\mathbf{H}}+\sigma_{e}\widetilde{\mathbf{H}},
\end{align}
where $\widehat{\mathbf{H}}=[\widehat{\mathbf{h}}_{1}, \widehat{\mathbf{h}}_{2}]$ is the channel estimate at the transmitter with independent and identically distributed (i.i.d.) elements $\widehat{h}_{j,k}\sim\mathcal{CN}(0,1)$, $\widetilde{\mathbf{H}}=[\widetilde{\mathbf{h}}_{1}, \widetilde{\mathbf{h}}_{2}]$ is the channel estimate error with i.i.d. elements $\widetilde{h}_{j,k}\sim\mathcal{CN}(0,1)$, for $1\leq j \leq n_{t}$ and, $\widehat{h}_{j,k}$ and $\widetilde{h}_{j,k}$ are the elements of the $j$th row and $k$th column of $\widehat{\mathbf{H}}$ and $\widetilde{\mathbf{H}}$, respectively, for \mbox{$k=1,2$}. The error variance is $\sigma_{e}^{2}=P^{-\alpha}$, where $\alpha$ is named as the CSIT quality scaling factor. 

We investigate the achievable throughput of RSMA, SDMA and NOMA with varying average SNR levels measured at the receivers. Let $S^{(l)}$ denote the number of channel uses in the $l$-th Monte-Carlo realization and $D_{s,k}^{(l)}$ denote the number of successfully recovered information bits by user-$k$ in the common stream (excluding the common part of the message intended for the other user) and its private stream. Then, we calculate the throughput as  
\vspace{-0.3cm}
\begin{align}
\mathrm{Throughput[bps/Hz]}=\frac{\sum_{l}(D_{s,1}^{(l)}+D_{s,2}^{(l)})}{\sum_{l}S^{(l)}}.
\end{align}

Figures~\ref{fig:tp_alpha06_noQoS}~and~\ref{fig:tp_alpha06_QoS} show the Shannon bounds and throughput levels achieved by RSMA, SDMA, and NOMA for $\alpha=0.6$ with and without QoS constraints for the users, respectively. The QoS constraint in Fig.~\ref{fig:tp_alpha06_QoS} is set as $R_{0}=0.1$~bps/Hz. Assigning higher values to $R_{0}$ frequently results in infeasible constraints in the precoder optimization problem for SDMA. 
The performance results clearly demonstrate that RSMA achieves a significant throughput gain over SDMA and NOMA. It can be observed that the performances of RSMA and NOMA are not affected by the QoS constraint, while the performance of SDMA degrades, even under a QoS constraint of $0.1$~bps/Hz. 

\section{Conclusion}
In this paper, we discussed the emerging problems and core services of 5G and KPIs and characteristics of 6G and how RSMA fits into these standards as an enabling technology. The literature and results so far imply that RSMA is a promising candidate in numerous application areas and use cases, and that RSMA can be used as a fundamental building block of the physical layer and lower MAC of 6G communications. There are still numerous open research topics for RSMA and addressing them will provide a better picture on the role and benefits of RSMA in 6G.

\vspace{12pt}
\color{red}

\end{document}